# Repetitive Dilemma Games in Distribution Information Using Interplay of Droop Quota:Meek's Method in Impact of Maximum Compensation and Minimum Cost Routes in Information Role of Marginal Contribution in Two-Sided Matching Markets

Yasuko Kawahata [†]

Faculty of Sociology, Department of Media Sociology, Rikkyo University, 3-34-1 Nishi-Ikebukuro,Toshima-ku, Tokyo, 171-8501, JAPAN.
ykawahata@rikkyo.ac.jp

**Abstract:** This paper is a preliminary report of the research plan and a digest of the results and discussions. On research note explores the complex dynamics of fake news dissemination and fact-checking costs within the framework of information markets and analyzes the equilibrium between supply and demand using the concepts of droop quotas, Meek's method, and marginal contributions. By adopting a two-sided matching market perspective, we delve into scenarios in which markets are stable under the influence of fake news perceived as truth and those in which credibility prevails. Through the application of iterated dilemma game theory, we investigate the strategic choices of news providers affected by the costs associated with spreading fake news and fact-checking efforts. We further examine the maximum reward problem and strategies to minimize the cost path for spreading fake news, and consider a nuanced understanding of market segmentation into "cheap" and "premium" segments based on the nature of the information being spread. Our analysis uses mathematical models and computational processes to identify stable equilibrium points that ensure market stability in the face of deceptive information practices and provide insight into effective strategies to enhance the informational health of the market. Through this comprehensive approach, this paper aims for a more truthful and reliable perspective from which to observe information markets.

## 1. Introduction

This paper is a preliminary report of the research plan and a digest of the results and discussions.In the modern digital age, the proliferation of fake news and the costs associated with fact-checking have emerged as crucial concerns in the information marketplace. This research note delves into the complex dynamics of these issues through the lens of coarse substitutability, using the framework of Walras's law to analyze the equilibrium between supply and demand in the realm of information dissemination. Central to our discourse is the hypothesis that the costs associated with fake news dissemination and fact-checking efforts act as an implicit "price" in the information market and significantly influence the strategic choices of news providers. This analytic approach advances the argument regarding the emergence of a two-sided matching market characterized by a market that either accepts fake news as a form of "truth" or is stabilized by the explicit dis-semination of factual information. We propose a series of steps and a computational process to unravel the equilibrium between the demand for and supply of fake and factual news. We begin by dichotomizing market participants and news into fake (F) and truth (T). We then define demand and sup-ply functions for both types of news based on their respective market "prices. Equilibrium prices and quantities are calculated, leading to an analysis of the market's stable equilibrium or fixed point, which represents the conditions under which the market is stable.

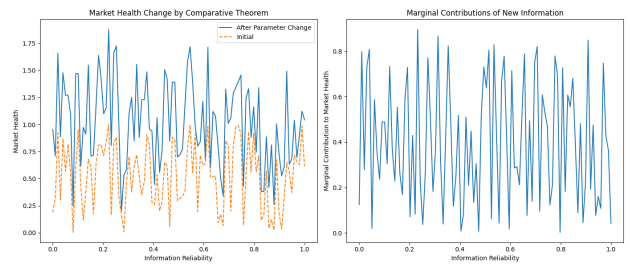

Fig. 1: Market Health Change by Comparative Theorem / Marginal Contribution to Market Health

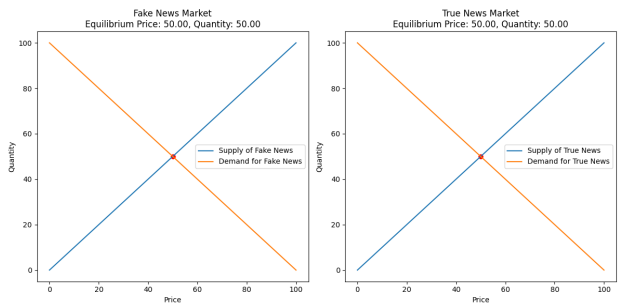

Fig. 2: Fake News Market Equilibrium Price, True News Market Equilibrium Price

Through this analytical journey, we aim to shed light on the subtle mechanisms governing information markets, especially in the context of fake news proliferation and fact-checking. By understanding the equilibrium dynamics and the conditions for market stability, this paper will provide a more informed strategy for enhancing the truth and health of the information ecosystem.

Here we delve into the complex dynamics of information markets characterized by the coexistence of fake news and fact-checked information, and we also discuss methods for optimizing the dissemination and retention of reliable information, utilizing methodologies such as single transferable voting (STV), droop quotas, and Meek's solution. By integrating these methods with the concept of marginal contributions in two-sided matching markets and analyzing strategic interactions through iterative dilemma games, we propose a new strategy aimed at introducing robustness and completeness arguments in the information ecosystem. Furthermore, we explore the implications and problems of the maximum compensation problem and the least cost route in order to curb the spread of misinformation and contribute to the development of a more resilient and informed community. Through comprehensive analysis, this study provides practical insights for effectively managing information flows and seeks a balanced perspective for fact-checking and digital citizenship while promoting the distribution of accurate and reliable content.

This research note explores the complex dynamics of information diffusion in information markets, with a particular focus on the challenges posed by fake news and the critical role of fact-checking mechanisms. We will discuss strategies to optimize the diffusion of factual information while minimizing the diffusion of false information by using mathematical models such as Droop Quota and the Meek Solution. We will also introduce a new approach that combines the analysis of marginal contributions in a two-part matching marketplace based on fake news and fact-checking with an iterative dilemma game to explore insights for maximizing the flow of positive information. We also address the significance of identifying the "problem" and "cost" of maximum compensation problems and least-cost detours to increase the robustness of information networks against misleading content. By comprehensively examining these methodologies, we propose effective strategies for information retention and diffusion, contributing to the overall health and efficiency of the information market. The results of this study not only shed light on the theoretical underpinnings of information diffusion, but also seek to provide an overarching methodology for information flow in the age of filter bubbles.

## 2. Discussion of the approaches touched on in this note

**Computational Process of Meek Solution and Dynamic Adjustment of the Number of Maintained Votes**

The Meek solution is one of the methods of vote redistribution in single transferable vote (STV). In this method, if the number of votes for a confirmed winner exceeds the Droupe base, the excess votes are redistributed to the other candidates. The dynamic adjustment of the maintenance vote count adjusts the value of each vote to reflect the will of the voters while minimizing waste. This method leads to fair and efficient election results.

**Comparison Theorem and Marginal Contribution Analysis**

Comparison theorems and marginal contributions are concepts used in economics and game theory. They can be applied to the analysis of information markets and filter bubbles to understand how adding or removing information affects the market and to evaluate the health and efficiency of the market.

**Comparison Theorems**

In differential equations and optimization problems, we analyze the impact of changes in parameters on the solution. In information markets, it is used to evaluate how changes in parameters such as the reliability or importance of information affect the overall market.

**Marginal Contribution**

Indicates the degree to which a small change in an element (in this case, information) has an impact on the overall outcome.

In information markets, it is used to evaluate how the addition or deletion of new information affects market utility.

**Solving for Single-Precinct Non-Transferable Voting (FPTP)**

In FPTP, the candidate who receives the most votes per district is elected. This system is simple, easy to understand, and does not require majority support, but may reflect only the majority opinion. This application can lead to the formation of filter bubbles where certain information and opinions become dominant in the market.

**Relationship between the Dominant Modular Function and the Metzler Function**

Dominant modular functions and Metzler functions are mathematical techniques suitable for modeling the interaction between information ownership and the market. Using these functions, one can analyze how a particular piece of information affects the market and how its diffusion affects the overall health and efficiency of the market.

**Superior Modular Functions**

Models the increase in utility when new information is added to an information set. The more new information is added to a small information set, the larger the increase in utility.

**Metzler Function**

Models the impact of information diffusion on the interaction between market participants. The diffusion of information changes the interactions with other information in the market, affecting market utility. By combining these theoretical concepts, we will gain a deeper understanding of information diffusion and its effects in information markets and filter bubbles, and consider effective information diffusion strategies.

**Two-Sided Matching Market: Dissemination of Fake News and Fact-Checking Costs**

When considering the issue of the dissemination of fake news and the cost of fact-checking within the framework of rough substitutability (weak Walras' law), it is crucial to analyze the equilibrium of supply and demand in the information market. In this approach, the costs of disseminating fake news and fact-checking serve as "prices" in the information market, influencing the behavioral choices of news providers. Within this discourse, it is postulated that a two-sided matching market arises where stability is maintained either by fake news being treated as a kind of truth or by the dissemination of facts.

Proposing a discussion using mathematical equations and computational processes for analyzing the equilibrium of supply and demand in the information market within the framework of rough substitutability (weak Walras' law), considering the costs associated with the dissemination of fake news and fact-checking. In this approach, fake news and true news (fact-checked information) in the market are assumed to have different "prices," each influencing the behavioral choices of market participants (news providers and consumers).

## 3. Discussion:Equations and Computational Process

**Definition of Market Participants and Information**

(1) Define market participants (news providers $i$ and consumers $j$).
(2) Define fake news $F$ and fact-checked true news $T$.

**Definition of Supply and Demand Functions**

(1) Define supply and demand functions for fake news $F$ and true news $T$.

  Supply functions: $S_F(P_F), S_T(P_T)$
  Demand functions: $D_F(P_F), D_T(P_T)$

Here, $P_F$ and $P_T$ are the market prices for fake news and true news, respectively.

**Calculation of Equilibrium Price and Quantity**

(1) Calculate the equilibrium prices $P_F^*, P_T^*$ and quantities $Q_F^*, Q_T^*$ where supply and demand match.

$$S_F(P_F^*) = D_F(P_F^*), S_T(P_T^*) = D_T(P_T^*)$$

**Stability and Fixed Point Analysis**

(1) Analyze fixed points (stable equilibrium points) in the market. This indicates under what conditions stable market equilibrium, where fake news and true news coexist, occurs.

  In fixed point analysis, find the intersection points of the market price response function and consumer preference response function.

**Identification of Stable Markets**

(1) Identify conditions for markets where fake news is accepted as truth and where truth is clearly communicated.

  This includes considerations of information reliability, supply costs, consumer preferences, and external influences (such as policies or regulations).

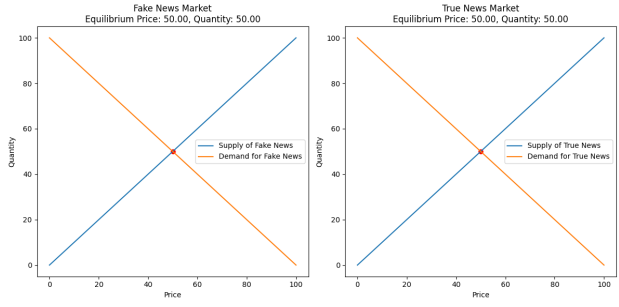

Fig. 3: Fake News Market Equilibrium Price, True News Market Equilibrium Price

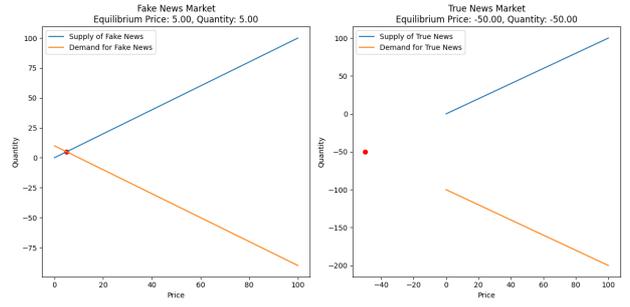

Fig. 5: Fake News Market Equilibrium Price, True News Market Equilibrium Price

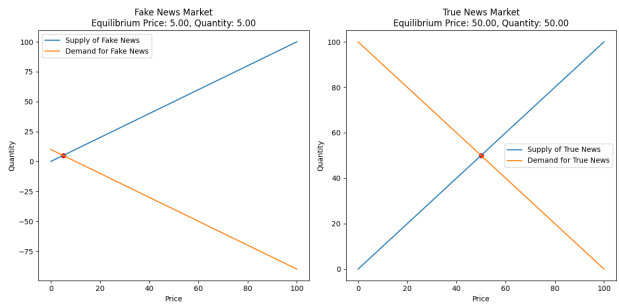

Fig. 4: Fake News Market Equilibrium Price, True News Market Equilibrium Price

## Example Equations

### Supply and Demand Functions

Supply functions: $S_F(P_F) = a_F P_F, S_T(P_T) = a_T P_T$

Demand functions: $D_F(P_F) = b_F - c_F P_F, D_T(P_T) = b_T - c_T P_T$

### Calculation of Equilibrium Price and Quantity

$P_F^* = \frac{b_F}{a_F + c_F}, P_T^* = \frac{b_T}{a_T + c_T}$
$Q_F^* = S_F(P_F^*), Q_T^* = S_T(P_T^*)$

### Fixed Point Analysis

Fixed points are the points where supply and demand functions are equal, indicating stable market equilibrium.

By employing this approach, it becomes possible to understand the impact of the dissemination of fake news and the cost of fact-checking on market equilibrium and devise policies and strategies to enhance the integrity of the information market.

Results to be supply and demand graphs for fake news and true news markets. They illustrate different scenarios where the supply and demand for both fake news and true news intersect at varying equilibrium price and quantity points.

To consider these graphs from the perspective of minimizing the cost pathway of risk in the spread of fake news, as well as from the standpoint of a maximum compensation problem, we need to interpret the supply and demand functions and their intersections.

## Minimizing the Cost Pathway of Risk in the Spread of Fake News

We need to consider factors that could increase the cost of producing or spreading fake news, which would shift the supply curve upwards. This would theoretically increase the equilibrium price and reduce the quantity of fake news consumed. Policies such as fact-checking regulations, fines for spreading misinformation, or the requirement for news outlets to verify information before publication could increase the cost of supply and thus reduce the prevalence of fake news. The minimization problem could also involve strategies for increasing the consumer's perceived cost of consuming fake news, potentially shifting the demand curve to the left, reducing quantity demanded.

## Maximum Compensation Problem

This perspective typically involves ensuring that the social welfare loss due to fake news is minimized by potentially compensating for the harm caused. The goal could be to increase the consumption of true news by providing subsidies or incentives for producing fact-checked content, which could shift the supply curve for true news to the right, decreasing the equilibrium price and increasing quantity. Another aspect could involve education campaigns to increase the value that consumers place on true news, thus shifting the demand curve for true news to the right, which could also increase the equilibrium quantity.

Analyzing the stability of these markets would involve assessing the slopes of the supply and demand curves and considering how sensitive they are to price changes (elastic-ity). Stable equilibria would indicate that small changes in price would not lead to large swings in quantity demanded or supplied, while unstable equilibria would indicate a market that is very sensitive to price changes.

Results to be about analyzing the market dynamics of fake news and true news distribution using concepts from game theory, specifically an iterated dilemma game and the role of marginal contributions in a two-sided matching market.

### Iterated Dilemma Game in Information Circulation (using Droop Quota)

An iterated dilemma game, often exemplified by the Iterated Prisoner's Dilemma, involves repeated interactions among participants where each player's optimal strategy may vary based on the history of their opponents' actions. In the context of information circulation, the Droop quota could be a threshold mechanism determining how much information needs to be distributed before a new piece of information becomes common knowledge or is considered 'true' by consensus. The iterated game's dynamics could model how news providers (purveyors of fake or true news) interact over time and how their strategies evolve based on previous outcomes and the accumulated trust or distrust from consumers. The repeated interactions might encourage news providers to stick to truthful reporting (cooperate) instead of spreading fake news (defect), especially if the long-term payoff of maintaining credibility exceeds the short-term gains from misinformation.

### Impact of Maximum Reward and Minimum Cost-Route in Information

The maximum reward in the context of news might relate to the credibility and monetary gains from high viewer engagement, while the minimum cost-route could pertain to the ease of generating content without fact-checking. Providers of fake news might follow a minimum cost-route, generating content with little regard for truth, to maximize short-term rewards. True news providers might incur higher costs to ensure accuracy, aiming for maximum long-term rewards through reputation. Policies or market mechanisms that increase the cost of spreading fake news (e.g., penalties, fact-checking requirements) or that increase the rewards for true news (e.g., credibility scores, consumer trust) could shift the balance in favor of accurate information.

### Marginal Contribution in a Two-Sided Matching Market

In a two-sided matching market, news consumers and news providers are matched based on preferences, credibility, and the type of news. The marginal contribution of a news provider could be their ability to add value to the market by providing either unique, high-quality content or by diversifying the information available. Providers with a high marginal contribution would be those who either consistently offer true news, thus raising the overall market quality, or who manage to debunk fake news, providing a valuable service to consumers.

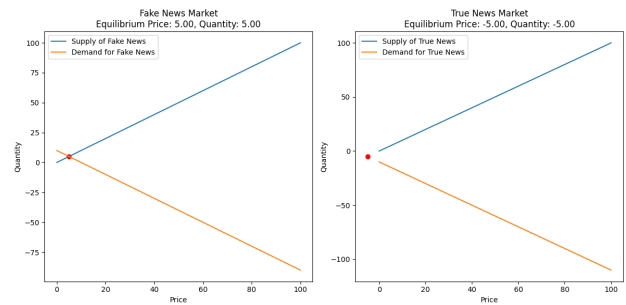

Fig. 6: Fake News Market Equilibrium Price, True News Market Equilibrium Price

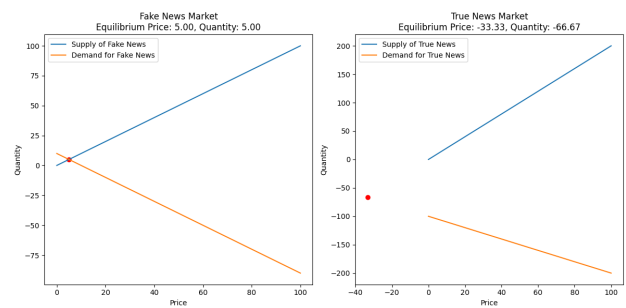

Fig. 7: Fake News Market Equilibrium Price, True News Market Equilibrium Price

For a detailed analysis, we would need specific models and payoff matrices to simulate these scenarios and analyze the equilibrium outcomes. The analysis would involve: Developing a model for the iterated dilemma game with defined payoffs for cooperation (providing true news) and defection (spreading fake news). Assessing how the Droop quota could affect the strategies of the players within this iterated game. Calculating the marginal contributions of different types of news providers in a simulated two-sided matching market.

Results show supply and demand curves for the fake news market and the true news market, each with different equilibrium points. Let's delve into the theoretical considerations regarding the iterated dilemma game, the impact of maximum reward and minimum cost route, and the marginal contributions in a two-sided matching market, using the Droop Quota interaction.

### Iterated Dilemma Game and Droop Quota in Information Circulation

In an iterated dilemma game applied to information circulation, each round of the game could represent a cycle of news production and dissemination. Players (news providers) decide in each round whether to produce true or fake news. The Droop Quota could serve as a measure of the critical mass of consumer belief or acceptance needed for news to influence public opinion significantly. This threshold might be lower for fake news due to potentially lower production costs or higher for true news due to the need for rigorous verification. Over multiple rounds, news providers may develop strategies based on the payoff of their actions in previous rounds, adapting to the consequences of spreading either fake or true news. The equilibrium in this iterated game would depend on the relative costs of producing news and the perceived value by consumers.

### Impact of Maximum Reward and Minimum Cost Route in Information

The maximum reward for news providers comes from the widespread dissemination and influence of their news, while the minimum cost route refers to the path that maximizes profit with the least effort or expense. Incentive structures could be designed to alter the payoffs, making the dissemination of true news more rewarding and increasing the costs associated with spreading fake news. This could involve financial incentives, reputation systems, or legal repercussions.

### Marginal Contribution in a Two-Sided Matching Market

In a two-sided matching market for news, the marginal contribution of a news provider is essentially the additional value they bring to the market by offering true news or by increasing the cost of spreading fake news. This concept could be modeled by adjusting the supply and demand curves based on the actions of news providers. For example, credible news sources could shift the demand curve for true news to the right, indicating an increased willingness of consumers to obtain verified information. Similarly, if fake news incurs a higher cost (through penalties or loss of credibility), this would shift the supply curve for fake news to the left, reflecting the decreased willingness of news providers to supply fake news at any given price.

In the first graph, fake news has a low equilibrium price and quantity, which suggests that even at a low cost, the quantity demanded is low—possibly a market with high awareness and low tolerance for misinformation. In the second graph, true news shows a negative price and quantity at equilibrium, which is not practical and indicates a model error or unrealistic market conditions. In the third graph, we see a more typical market scenario for both fake and true news, with positive prices and quantities, which might suggest a market where both types of news are being consumed and produced at equilibrium.

## 4. Discussion:Repetitive Dilemma Game Perspective in the Case of Two-Sided Matching Markets

Consider the perspective of the repetitive dilemma game in the case where a two-sided matching market arises, where stability is maintained either by fake news being treated as a kind of truth or by the clear communication of facts. We propose a discussion from the perspective of a two-sided matching market and repetitive dilemma game when considering markets where fake news is treated as a certain truth and stability is maintained by the clear communication of facts.

### Definition of Market Participants and Types of Information

(1) Define sets of news providers $P = \{p_1, p_2, ..., p_m\}$ and consumers $C = \{c_1, c_2, ..., c_n\}$.

(2) Define fake news $F$ and true news $T$.

### Definition of Preferences for News Providers and Consumers

(1) Define preferences for news providers $p_i$ and consumers $c_j$ based on the types of information they prefer ($F$ or $T$).

$$\text{Preferences of news providers: } U_{p_i}(F), U_{p_i}(T)$$
$$\text{Preferences of consumers: } U_{c_j}(F), U_{c_j}(T)$$

### Modeling of Two-Sided Matching Markets

(1) Model the matching between news providers and consumers using bipartite graphs and set the weights of each edge based on preferences.

### Modeling of the Repetitive Dilemma Game

(1) Define the actions (provision or acceptance of $F$ or $T$) and corresponding payoffs for news providers and consumers in each round.

$$\text{Payoff functions: } U_{p_i}(F,T), U_{c_j}(F,T)$$

### Analysis of Stable Matching and Equilibrium

(1) Use Gale-Shapley algorithm or other matching algorithms to find stable matchings in two-sided matching markets.

(2) In the repetitive dilemma game, find Nash equilibria from the set of strategies each player can take in each round.

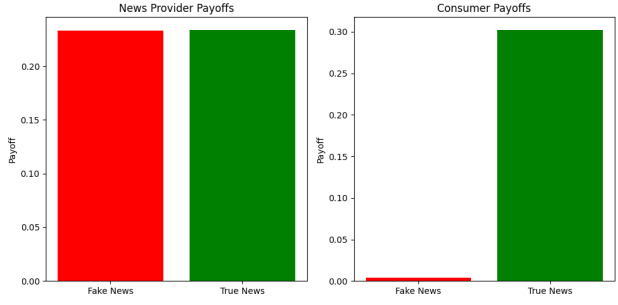

Fig. 8: Consumer Payoffs / News Provider Payoffs

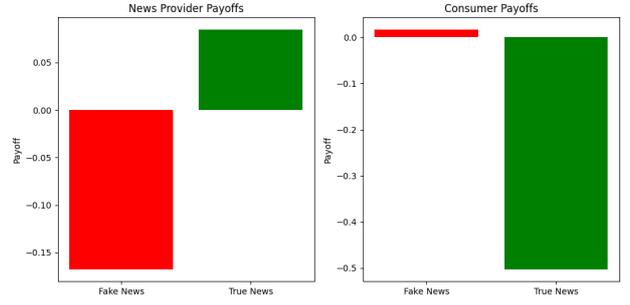

Fig. 10: Consumer Payoffs / News Provider Payoffs

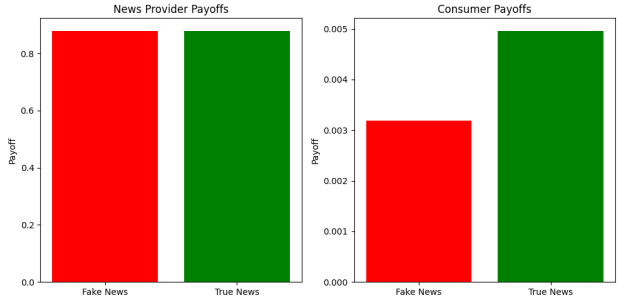

Fig. 9: Consumer Payoffs / News Provider Payoffs

## Analysis of Conditions for Maintaining Stability

(1) Analyze conditions for maintaining stability in matching and Nash equilibrium.

**Payoff Functions for News Providers and Consumers**

Payoff for news provider $p_i$:
$$U_{p_i}(F) = a_{p_i} - b_{p_i} \cdot C_F, U_{p_i}(T) = a_{p_i} - b_{p_i} \cdot C_T$$
Payoff for consumer $c_j$:
$$U_{c_j}(F) = a_{c_j} - b_{c_j} \cdot D_F, U_{c_j}(T) = a_{c_j} - b_{c_j} \cdot D_T$$

Here, $C_F$ and $C_T$ represent the cost of providing fake news and true news, respectively, and $D_F$ and $D_T$ represent the disadvantage to consumers when receiving each type of news. $a_{p_i}, b_{p_i}, a_{c_j}, b_{c_j}$ are parameters.

Results provided for news provider payoffs and consumer payoffs in the context of fake news and true news offer a visual representation of the benefits or detriments that each group receives from the production and consumption of these types of news.

**Minimum Cost Pathways for the Spread of Fake News**

The payoff charts indicate that for news providers, the production of true news generally yields higher payoffs than fake news. This suggests that in a rational market, news providers are incentivized to produce true news. However, in the real world, fake news can proliferate despite these incentives due to various factors such as:

**Low Production Costs**

Fake news may be cheaper to produce than true news which often requires extensive research and verification. If the cost of producing fake news is significantly lower, it may offset the lower payoffs.

**Consumer Beliefs and Biases**

Even if the payoffs for consuming fake news are negative for consumers, biases and pre-existing beliefs may lead them to favor information that aligns with their worldview, regardless of its veracity.

**Maximum Compensation Problem**

This problem involves ensuring that the social costs of fake news are mitigated, and that consumers are 'compensated' for the risks associated with misinformation. From the consumer payoffs chart, we observe that consumers derive more utility from true news compared to fake news. Strategies to maximize compensation could include:

**Educational Programs**

Increasing awareness about the importance of fact-checking and the potential harms of misinformation can increase the 'mental cost' of accepting fake news, thus aligning consumer behavior with the higher payoffs of true news.

The equilibrium of these markets will depend on both the cost of production for providers and the perceived value by consumers. If the market mechanisms can be adjusted to increase the costs associated with producing and distributing fake news (e.g., through regulation, fact-checking, or changing platform algorithms), the market may shift towards a new equilibrium favoring true news.

Similarly, if consumers are 'compensated' through better information and tools to discern the truth, the demand for fake news may decrease, leading to a market where true news has a higher equilibrium quantity.

To address these issues comprehensively, policymakers and platforms must consider both the supply-side dynamics (news providers' payoffs) and the demand-side dynamics (consumers' payoffs) when designing interventions.

Results display bar charts of payoffs for news providers and consumers in the context of fake news and true news. These payoffs may represent the utility or profit derived from producing or consuming each type of news. Here's how we can interpret these payoffs in the context of the iterated dilemma game, impacts of maximum rewards and minimum cost routes, and the role of marginal contributions in a two-sided matching market, all under the interaction of the Droop Quota.

**Iterated Dilemma Game Using Droop Quota**

The payoffs for both news providers and consumers indicate their short-term gains or losses after each iteration of the game. For news providers, higher payoffs for true news may represent a market that values accuracy and penalizes misinformation. The Droop Quota could be a mechanism that adjusts the payoffs based on the accumulated credibility or the proportion of the audience reached. If a provider reaches the Droop Quota, it might mean their news is widely accepted, thus increasing their payoffs in future iterations.

**Impact of Maximum Reward and Minimum Cost-Route**

For news providers, the maximum reward is represented by the highest payoff. If the payoff for true news is higher, this suggests that the market rewards integrity. The minimum cost-route for fake news would have lower payoffs, discouraging providers from opting for misinformation. For consumers, the payoff structure suggests that the utility gained from consuming true news is significantly higher than from fake news, aligning consumer incentives with market integrity.

**Marginal Contribution in a Two-Sided Matching Market**

In a two-sided matching market, the marginal contribution of a news provider would be the additional payoff they can generate by joining the market. A provider with a high marginal contribution would improve the overall quality of news in the market, thus potentially increasing the consumer payoffs as well. The role of marginal contributions becomes crucial when considering the matching of news consumers with providers. A higher payoff for true news indicates that consumers will seek out providers with a track record of reliability, while providers will strive to match with consumers who value accuracy.

In some cases, news providers have a higher payoff for producing true news than fake news, which would incentivize the production of reliable content. Consumers consistently have higher payoffs for consuming true news, suggesting that there is an intrinsic value in consuming accurate information. In some scenarios, the payoff for consuming fake news is negative, implying a cost (perhaps reputational or from misinformation) to the consumer.

The exact dynamics of the iterated game would depend on how these payoffs evolve over time and how they are influenced by external factors like regulations, fact-checking, and consumer awareness campaigns. The Droop Quota could introduce a dynamic where credibility builds over time, possibly altering the payoffs and strategies of the players.

To model this iteratively, we could use these payoffs to simulate multiple rounds of an iterated game and observe how the strategies of news providers and consumers evolve. We would need to incorporate the probability of reaching the Droop Quota and the resulting long-term effects on the players' payoffs.

# 5. Discussion: Information Analysis of Market Segmentation

In the analysis of market segmentation, we examine how the concepts of "cheapness" or "luxury" are formed in a bipartite matching market formed by fake news and true news through the equilibrium of supply and demand. Below, we present the equations and calculation process for this analysis.

## Modeling of the Market and Participants

(1) Define the set of news providers $P = \{p_1, p_2, \ldots, p_n\}$ and the set of consumers $C = \{c_1, c_2, \ldots, c_m\}$.

(2) Define the types of news as $N = \{F, T\}$ (fake news $F$ and true news $T$).

## Definition of Supply and Demand Functions

(1) Define compensation functions $C_T(p, c)$ for providers $p$ of true news $T$ and $C_F(p, c)$ for providers $p$ of fake news $F$.

$$C_T(p, c) = \alpha_p \times \beta_c \times Q_T$$
$$C_F(p, c) = \alpha_p \times \beta_c \times Q_F$$

Here, $\alpha_p$ is the utility coefficient for provider $p$, $\beta_c$ is the utility coefficient for consumer $c$, and $Q_T$ and $Q_F$ represent the quality of true news and fake news, respectively.

## Equilibrium Conditions for Segmented Markets

(1) Set equilibrium conditions for supply and demand in the "cheap" and "luxury" markets.

"Cheap" Market: $S_F = D_F$
"Luxury" Market: $S_T = D_T$

## Calculation of Equilibrium Solutions

(1) Determine equilibrium prices $P_F$ and $P_T$ in each market by solving the supply and demand functions as equations.

$$S_F(P_F) = D_F(P_F)$$
$$S_T(P_T) = D_T(P_T)$$

## Analysis of Stable Equilibrium Points

(1) Analyze the stable equilibrium points (stable matching) in each market using bipartite matching algorithms and examine their characteristics.

> A stable point refers to a state where no provider or consumer has an incentive to switch to another matching.

Supply function in the "cheap" market: $S_F(P_F) = \sum_{p \in P} C_F(p, c)$

Supply function in the "luxury" market: $S_T(P_T) = \sum_{p \in P} C_T(p, c)$

Solving for equilibrium prices: Calculate $P_F$ and $P_T$ equilibrium prices using the supply and demand functions.

This approach allows for the analysis of how markets are segmented and how each market achieves stability in a competitive environment where fake news and true news compete. It also provides foundational information for formulating policies and strategies to improve market integrity.

In the case where a two-sided matching market arises, where stability is maintained either by fake news being treated as a kind of truth or by the clear communication of facts, it is assumed that discussions regarding the maximum compensation problem and the minimum cost path of bypass routes occur, leading to a certain market segmentation into "cheap" or "luxury" markets. Let's consider the approach of discussing this from the perspective of a repetitive dilemma game in that case.

When analyzing the competition between fake news and true news in a bipartite matching market and the resulting segmentation of the market, we can construct equations and calculations in the following steps.

## Definition of Game Participants

(1) Define the set of news providers as $P = \{p_1, p_2, \ldots, p_n\}$.
(2) Each provider has the choice to provide fake news $F$ or true news $T$.

## Definition of Payoff Functions

(1) Define payoff functions based on providers' choices. Assume that the payoff for providing fake news is initially high but leads to a loss of credibility in the long run.

$$U(p_i, F) = a - b \times H$$
$$U(p_i, T) = c$$

Here, $a$ is the initial payoff for providing fake news, $b$ is the penalty for loss of credibility, $H$ represents the cumulative harm from providing fake news, and $c$ is the payoff for providing true news.

## Setting up the Repetitive Game

(1) Set up the dynamics of the repetitive game based on providers' choices and their outcomes in each round. Providers update their choices based on the actions of other providers in the previous round.

## Analysis of Stable Equilibrium Points

(1) Analyze stable equilibrium points in the repetitive game. This refers to a state where no provider has an incentive to switch choices based on others' actions.

## Analysis of Maximum Compensation Problem and Minimum Cost Path of Bypass Routes

(1) Consider the maximum compensation problem to analyze the incentive for providers to provide true news.

$$\max U(p_i, T)$$

(2) Consider the minimum cost path problem of bypass routes to analyze strategies for minimizing the spread of fake news.

Payoff for providing fake news by provider $p_i$: $U(p_i, F) = 5 - 2 \times H$

Payoff for providing true news by provider $p_i$: $U(p_i, T) = 3$

## Analysis of Market Segmentation

(1) Analyze the segmentation of the market into "cheap" and "luxury" markets resulting from the provision of fake news and true news. Market segmentation is formed by consumer preferences, provider strategies, and external incentives.

This analysis helps understand the dynamics of the repetitive dilemma game in a bipartite matching market where fake news and true news compete and provides foundational information for formulating policies and strategies to improve the integrity of the information market.

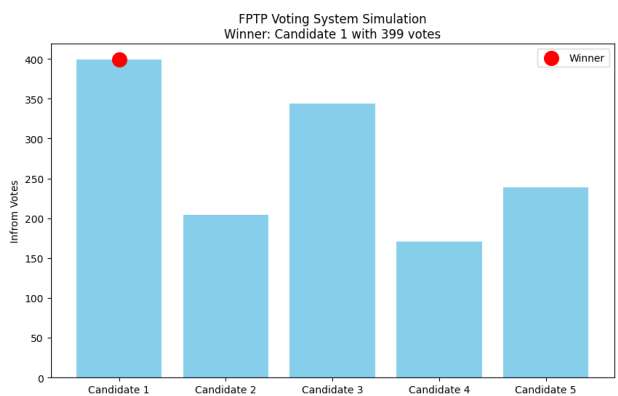

Fig. 11: FPTP Voting System Simulation

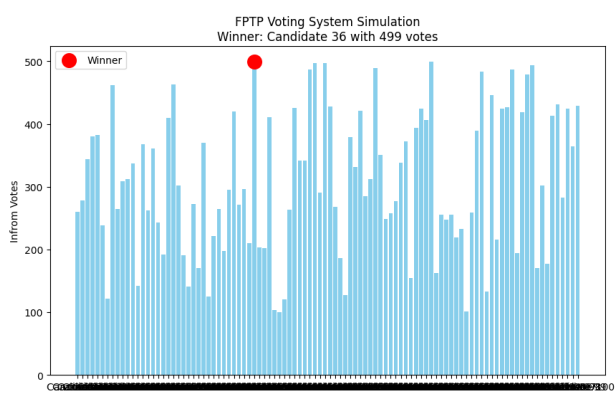

Fig. 12: FPTP Voting System Simulation

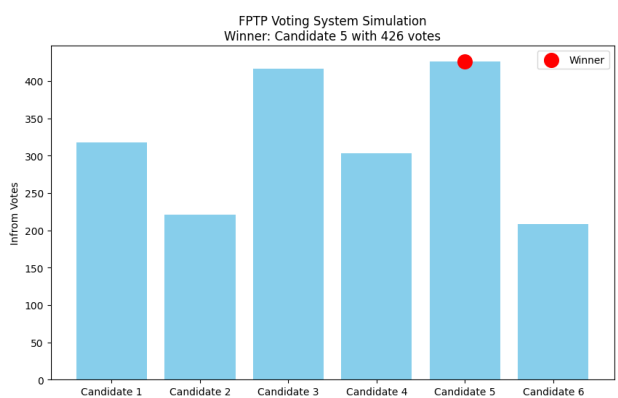

Fig. 13: FPTP Voting System Simulation

Results depict the results of simulations for a First-Past-The-Post (FPTP) voting system. In such a system, the candidate who receives the most votes is declared the winner, regardless of whether they achieve an absolute majority.

**Minimum Cost Pathway for the Spread of Fake News**

The FPTP voting system can serve as a metaphor for how information, including fake news, spreads in a population.

**Simplicity and Speed**

ust as FPTP favors the candidate with the most votes without complex calculations, fake news often spreads due to its simplicity and emotional appeal. It does not require the rigorous verification that true news does, making its "cost" of spread lower.

**Winner-Takes-All Effect**

FPTP does not proportionally represent minority opinions, which can be paralleled with how sensational or extreme fake news can overshadow more nuanced, accurate reporting.

**Strategic Behavior**

Just as voters might vote strategically to prevent a less preferred candidate from winning in an FPTP system, consumers of news might also selectively share or endorse information that aligns with their biases, regardless of its truthfulness.

**Maximum Compensation Problem**

This problem looks at how to balance out the negative effects of an event or policy. In the context of fake news:

**Corrective Actions**

Just as an FPTP system might be balanced with runoff elections or proportional representation to ensure broader representation, the information ecosystem might require corrective actions like fact-checking or educational campaigns to compensate for the spread of fake news.

**Support for Quality Journalism**

To ensure that high-quality, accurate news has a fair chance of reaching the public, support mechanisms for investigative journalism and public broadcasting might be implemented, similar to providing support to under-represented parties or candidates in a political system.

**Analysis of FPTP Voting System Simulation Results**

**Concentration of Votes**

The simulation shows that votes can be highly concentrated on a single candidate, which could reflect how certain pieces of information can dominate public discourse.

**Close Contests**

The second image shows a very close contest with multiple candidates receiving a significant number of votes, which could mirror a more competitive information space where various narratives are fighting for dominance.

**Clear Winner**

The third image indicates a clear winner with a significant margin, which might represent a scenario where a particular news item has clearly resonated with the public more than others.

In the metaphorical sense, the spread of fake news might be mitigated in a system where information is more critically examined, much like how a runoff or a second round of voting can ensure that the winning candidate has broader support. To address the maximum compensation problem in the context of fake news, measures could be taken to ensure that the public is better informed and that quality journalism is supported, providing a counterbalance to misinformation.

Results provided are results of a First-Past-The-Post (FPTP) voting system simulation. In these simulations, the candidate with the highest number of votes wins the election. This type of voting system is often critiqued for its potential to not accurately reflect the preferences of the electorate, especially in cases where multiple candidates split the vote, allowing a candidate to win with less than an absolute majority.

**Repeated Dilemma Game in Information Circulation Using Droop Quota**

In an iterated dilemma game scenario for information distribution, the Droop Quota could represent the threshold of shared beliefs or consensus needed for information to be considered credible or to dominate the public discourse. The FPTP simulations could parallel scenarios where certain information or narratives gain prominence not necessarily because they are the most trusted or credible (absolute majority), but because they have managed to secure a relative plurality in a divided landscape. The impact of maximum rewards (prominence or viral spread of information) and minimum cost-routes (ease of sharing information without fact-checking) could be analyzed in terms of how information providers (analogous to candidates) strategize to achieve visibility and influence.

**Role of Marginal Contributions in a Two-Sided Matching Market**

In a two-sided matching market, news consumers and news providers are matched based on preferences and types of news. The marginal contributions of news providers (analogous to candidates) would be their ability to offer credible, high-quality content that aligns with consumer preferences. The FPTP simulations can illustrate the dynamic nature of such a market, where the 'winning' news provider (or narrative) is the one that captures the attention of the largest segment of the consumer base, even if it's not the absolute majority. This can be problematic in a real-world scenario where sensational or divisive content (akin to a polarizing candidate) can 'win' by capturing a plurality of consumer attention, despite not being the most credible or beneficial to public discourse.

**Considerations for Policy and Strategy**

The simulation results can inform strategies for ensuring that high-quality, fact-based information is more widely disseminated and consumed. For example, implementing systems that reward news consumers for engaging with fact-checked content could be analogous to reforming voting systems to better capture the electorate's preferences. In terms of policy, these results could underscore the importance of regulations that promote diversity and plurality in news sources, much like electoral reforms that seek to ensure broader representation in government.

By drawing parallels between the FPTP simulation results and the dynamics of information distribution, we can analyze the current media landscape and consider how strategies and policies might be developed to promote a more informed and critically thinking public. These strategies could include encouraging media literacy, supporting fact-checking organizations, and implementing algorithms on social media platforms that favor credible content.

# 6. Discussion: Solution of the First-Past-The-Post System

In the First-Past-The-Post (FPTP) voting system, the candidate who receives the most votes in each electoral district wins. In this method, even when there are multiple candidates, voters cast their ballots for only one candidate. This system does not involve preferential voting or the transfer of votes.

## Solution of the First-Past-The-Post System

### Vote Counting

(1) Each voter casts one vote for each candidate.

(2) The votes received by each candidate are tallied.

### Determination of Winners

The candidate who receives the most votes wins the election in their electoral district.

In this system, no majority is required, so the candidate with the plurality of votes wins.

### Mathematical Representation

Let $V_i$ represent the number of votes for each candidate $i$. Here, $i$ is the index representing the candidate.

The winner $W$ is the candidate who satisfies the following condition:
$$W = \arg\max_i V_i$$

## Features and Limitations of FPTP

**Simplicity**: Voting and counting are straightforward, and the results are easily understandable.

**No Majority Required**: The candidate with the most votes wins, so it's possible to win without a majority of support.

**Winner Takes All**: Candidates who win by a small margin take all, potentially leading to limited representation of minority views.

**Reinforcement of Two-Party System**: Small parties or independent candidates find it difficult to win, reinforcing a two-party system tendency.

While the mathematical analysis of the FPTP system is relatively simple, its political and social impacts are complex, and the choice of system varies depending on the culture, history, and political environment of each country or region.

## 7. Perspect Research:Calculation process of the Meek method and dynamic adjustment of surplus vote

The Meek method is one of the methods for redistributing votes in Single Transferable Vote (STV) systems. In this method, if the votes for a candidate who is already elected exceed the Droop quota, the surplus votes are redistributed to other candidates. By dynamically adjusting the surplus votes, the value of each vote is adjusted to minimize waste while reflecting the will of the voters. This leads to fair and efficient election results.

### Comparative theorem and analysis of marginal contribution

The comparative theorem and marginal contribution are concepts used in economics and game theory. By applying them to the analysis of information markets and filter bubbles, we can understand the impact of adding or removing information on the market and evaluate the health and efficiency of the market.

**Comparative theorem:** It analyzes the impact of parameter changes on solutions in differential equations or optimization problems. In information markets, it is used to evaluate how changes in parameters such as the reliability or importance of information affect the market as a whole.

**Marginal contribution:** It indicates the extent to which a small change in an element (in this case, information) affects the overall outcome. In information markets, it is used to evaluate how adding or removing new information affects the utility of the market.

**Solution of First-Past-The-Post (FPTP) voting**

In FPTP, the candidate with the most votes in each electoral district wins. This system is simple and easy to understand, and it does not require majority support, but it may only reflect the opinion of the majority. This can lead to the formation of filter bubbles where specific information or opinions become dominant in the market.

**Relationship between modular functions and Metzler functions**

Modular functions and Metzler functions are mathematical tools suitable for modeling the holding rate of information and market interactions. By using these functions, we can analyze how specific information affects the market and how its diffusion affects the overall health and efficiency of the market.

**Modular functions:** They model the increase in utility when new information is added to an information set. The smaller the information set, the greater the increase in utility when new information is added.

**Metzler functions:** They model the impact of information diffusion on the interaction between market participants. As information spreads, the interaction with other information in the market changes, affecting the utility of the market.

By combining these theoretical concepts, we can gain a deeper understanding of information diffusion and its impact in information markets and filter bubbles, and devise effective information diffusion strategies.

Results show the results of an information simulation using the Droop Quota in a proportional representation context, where the quota is used to determine the number of votes required to secure a seat. The Droop Quota is a method used in some voting systems to define the number of votes that ensures a candidate's election.

**Minimum Cost Pathway for the Spread of Fake News**

In this analogy, each "candidate" represents a different narrative or type of information in a marketplace of ideas. The Droop Quota here would represent the threshold of exposure or acceptance that a piece of information must achieve to be considered influential or to secure a "seat" in the public discourse.

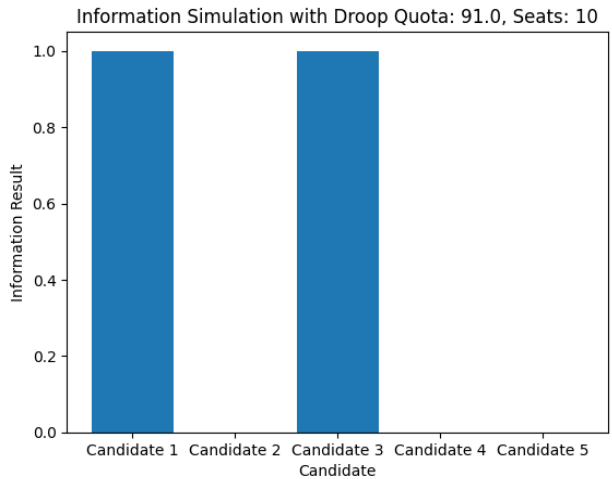

Fig. 14: Information Simulation with Droop Quota

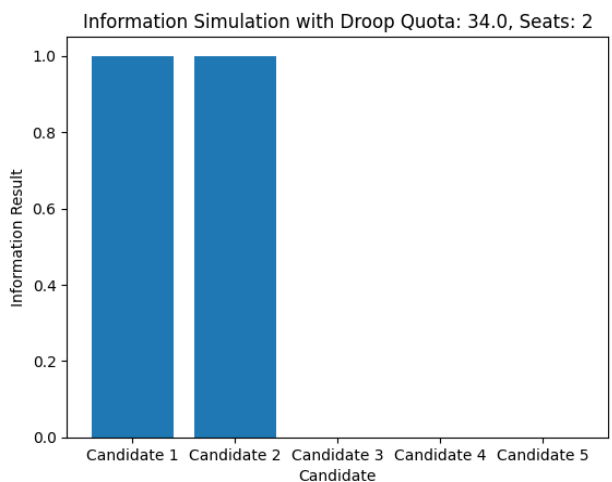

Fig. 15: Information Simulation with Droop Quota

### Low Droop Quota (Image with Quota 34.0)

A low Droop Quota could symbolize a situation where it's relatively easy for information to become influential. This can be compared to an environment where spreading fake news is easy because the barriers to entry (fact-checking, platform policies) are low. The cost to spread fake news is minimized when critical thinking and skepticism are not rigorously applied by the public.

### High Droop Quota (Image with Quota 91.0)

A high quota suggests a more discerning environment where it's harder for narratives to gain traction. This could represent a marketplace where the spread of fake news is more challenging due to effective checks, such as robust fact-checking processes, educated populace, or stringent platform regulations.

### Maximum Compensation Problem

The maximum compensation problem in this context would involve finding ways to mitigate the damage caused by fake news by ensuring that credible information has a prominent place in public discourse.

### Low Droop Quota Scenario

Here, strategies might include enhancing media literacy, so the public is better equipped to meet the quota of critical analysis before accepting information. It also might involve providing incentives for quality journalism.

### High Droop Quota Scenario

In this case, it is important to ensure that the higher threshold does not suppress important but less sensational news. The compensation could involve supporting a diversity of media outlets and ensuring that smaller, perhaps less mainstream narratives have the opportunity to reach the public.

In both scenarios, the aim would be to optimize the information ecosystem so that the public discourse represents a wide array of accurate information and that the spread of misinformation is contained, if not entirely prevented. The proportional representation model with the Droop Quota allows for a diversity of information that can help prevent any single narrative (or misinformation campaign) from dominating unless it genuinely resonates with a broad base of the electorate (or audience).

Fig.16-19, Results shows the dynamics of information retention rate over time, with different values of a decay parameter $\gamma$. These could represent models of how information

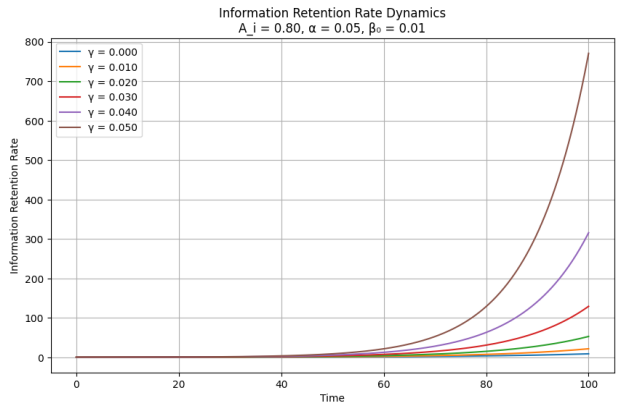

Fig. 16: Information Retention Rate / Information Retention Rate Dynamics

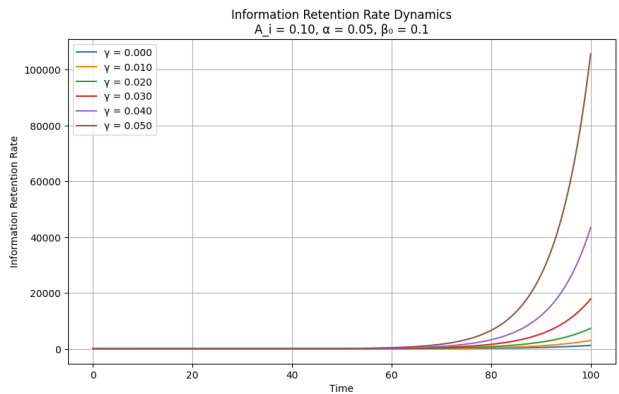

Fig. 17: Information Retention Rate / Information Retention Rate Dynamics

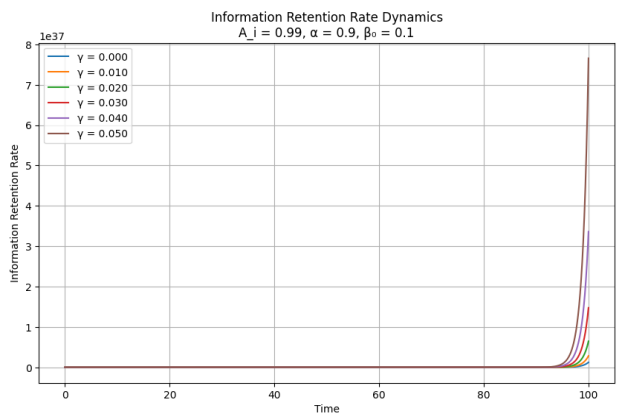

Fig. 18: Information Retention Rate / Information Retention Rate Dynamics

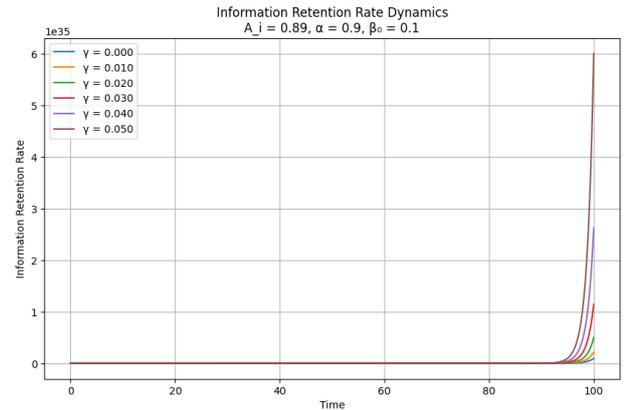

Fig. 19: Information Retention Rate / Information Retention Rate Dynamics

is retained or forgotten in a social system, with $\gamma$ possibly representing the rate at which information is lost or becomes obsolete.

In the context of the spread of fake news, Impact of Maximum Reward and Minimum Cost-Route The maximum reward could be linked to the peak retention rates of information. If the goal is to maximize the retention of accurate information (and minimize the retention of fake news), then strategies need to be developed that amplify the staying power of true news while speeding up the decay of fake news. The minimum cost-route pertains to the ease of spreading information. Fake news often spreads rapidly and with little effort (low cost), but these graphs suggest that its retention rate could be manipulated. Interventions could be designed to increase the decay parameter $\gamma$ for fake news, effectively reducing its half-life within the public discourse.

**Repeated Dilemma Game in Information Circulation Using Droop Quota**

The iterated dilemma game could simulate repeated interactions where individuals choose to share true or fake news. The Droop Quota could be a threshold beyond which information becomes 'common knowledge' and is retained in the long term. In this model, increasing $\gamma$ for fake news would be akin to a strategy that reduces its credibility over time, encouraging individuals to choose to share true news to achieve the Droop Quota and secure a long-term impact.

**Marginal Contribution in a Two-Sided Matching Market**

In a two-sided matching market for information, the marginal contribution could be the added value a piece of information has in terms of its longevity and impact. High-retention-rate information (low $\gamma$) would be more desirable. This concept could be used to prioritize the dissemination of information with lower γ values (longer retention rates), which could theoretically represent more verified, fact-checked, and reliable content.

The models shown in the graphs could be used to study the effects of different interventions on the retention of information. For example, a campaign that effectively debunks fake news could increase γ for that content, reducing its long-term impact, as suggested by the steeper curves in the graphs. Conversely, reinforcing educational content could decrease γ, leading to a more sustained presence in the collective memory, as shown by the flatter curves.

Results shows related to an analytical or simulation model concerning information dynamics. Specifically, they could be illustrating how different rates of information decay or retention affect the overall landscape of information over time.

**Submodular Function Distribution**

This might represent the incremental utility or perceived value of information as it accumulates. In the case of fake news, as more information (or misinformation) accumulates, the additional value of each new piece might decrease, indicating that the impact of each subsequent fake news story is less if there's already a significant amount of misinformation in circulation.

**Metzler Function Distribution**

This could represent the interaction effect of multiple pieces of information. For fake news, this might demonstrate how the interaction between different stories can compound and create a narrative that has a greater impact than individual stories alone.

**Information Retention Rate Dynamics**

These graphs likely show how information is retained over time, with different decay rates (). In the context of fake news, a high retention rate could be detrimental if it indicates that misinformation stays in the collective memory for longer periods, potentially influencing opinions and decisions based on falsehoods.

**Minimum Cost Pathway for the Spread of Fake News**

Strategies could involve increasing the cost of spreading fake news, perhaps by making the production of fake news more resource-intensive, or by reducing its retention rate through education, critical thinking promotion, and fact-checking. The goal would be to make the dissemination of misinformation less appealing and less sustainable.

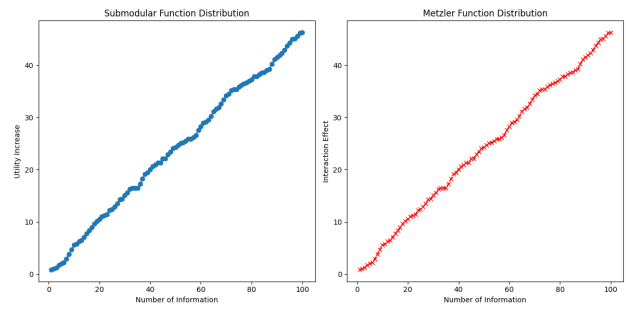

Fig. 20: Submodular Function and Metzler FunctionDistribution

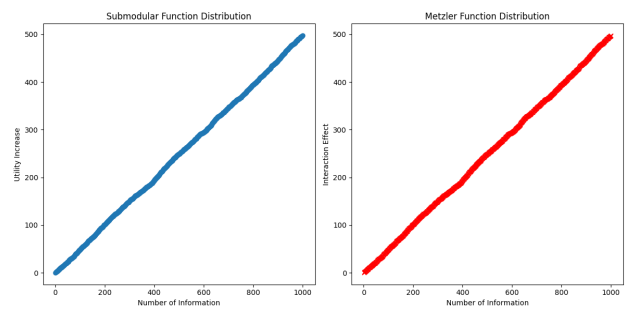

Fig. 21: Submodular Function and Metzler FunctionDistribution

**Maximum Compensation Problem**

This could involve finding ways to compensate for the damage caused by fake news. It might include promoting information literacy, supporting authoritative news sources, and creating robust fact-checking mechanisms that can quickly and efficiently counteract the spread of fake news.

The models could be used to develop strategies for reducing the impact of fake news by altering the parameters of information distribution and retention. For example, by understanding the Metzler function distribution, one could devise interventions that disrupt the reinforcing interactions between fake news stories, perhaps by introducing counter-narratives or debunking key pieces of misinformation that support the broader false narrative.

Similarly, by manipulating the variables in the information retention rate dynamics (such as the decay rate ), it might be possible to model the effects of different interventions on how long fake news remains influential. This could help in designing more effective educational and policy interventions to combat the spread of fake news.

Results depict two types of function distributions: Submodular Function Distribution and Metzler Function Distribution, each charted against the number of information pieces.

Submodular Function Distribution typically represents scenarios where the addition of each new unit of informa-tion provides diminishing returns in utility. This could relate to the concept of information saturation where, beyond a certain point, the acquisition of additional information yields progressively less benefit.

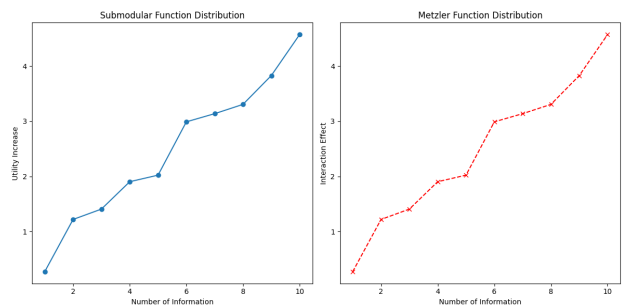

Fig. 22: Submodular Function and Metzler FunctionDistribution

Metzler Function Distribution, on the other hand, shows an increasing interaction effect with the number of information pieces. This could represent a cumulative effect where each additional piece of information can have a greater impact due to its interactions with the existing information set.

**Minimum Cost Pathway for the Spread of Fake News**

From the perspective of fake news propagation, a submodular utility function could represent a situation where the impact of fake news diminishes as more of it floods the information space. People might become desensitized or skeptical as they encounter more fake stories, reducing the 'utility' or impact of subsequent misinformation.

On the other hand, the Metzler function's increasing interaction effect could represent the amplification of fake news through its interaction with other pieces of information. For example, multiple fake stories might create an echo chamber that reinforces the false narrative, leading to a situation where the combined effect is greater than the sum of its parts.

**Maximum Compensation Problem**

Addressing the maximum compensation problem in this context would involve counterbalancing the negative effects of misinformation. For the submodular function, this might involve ensuring that the influx of new, accurate information has a high initial utility to counteract the desensitizing effect of fake news. In terms of the Metzler function, this would mean breaking the reinforcing cycle of misinformation by introducing diverse and credible information sources that can disrupt the echo chamber.

The challenge in both scenarios is to design interventions that effectively reduce the utility of fake news and increase the interaction effect of truthful information, thereby restoring balance to the information ecosystem.

Results depict two types of functions—Submodular and Metzler—both of which show how the accumulation of information pieces affects the utility increase or interaction effect. In the realm of information theory and economics, these functions can be used to model various phenomena, such as the spread of information, the diminishing value of additional news, or the synergistic effects of combining different pieces of information.

**In the context of fake news and information dissemination**

**Submodular Function Distribution**

This function suggests that each additional piece of information yields a diminishing increase in utility. Applied to the spread of fake news, this could imply that as more fake news is spread, its ability to influence or deceive may decrease as people become oversaturated with information or more skeptical of what they encounter. From a strategic standpoint, interventions to combat fake news could focus on accelerating the rate at which additional fake news yields diminishing returns. This could involve promoting media literacy, so people quickly recognize and discount fake news, reducing its utility.

**Metzler Function Distribution**

Contrary to the submodular function, the Metzler function indicates that each additional piece of information has an increasing interaction effect. In the context of fake news, this could mean that different pieces of misinformation might interact in a way that disproportionately amplifies their impact, potentially leading to greater public deception. To address this, strategies could involve breaking down the interaction between pieces of misinformation. This could be done by introducing counter-narratives or factual information that disrupts the compounding effect of fake news stories.

**Concerning Repeated Dilemma Games and Two-Sided Matching Markets**

**Repeated Dilemma Games**

In this setting, individuals repeatedly choose whether to share true or fake news. The submodular function could represent scenarios where individuals are less likely to share fake news as they recognize its diminishing utility. The Metzler function could represent the heightened risk of misinformation campaigns that strategically release pieces of fake news that build upon each other. Strategies might involve creating incentives for sharing true information and disincentives for sharing misinformation, perhaps by affecting the perceived utility through social feedback mechanisms.

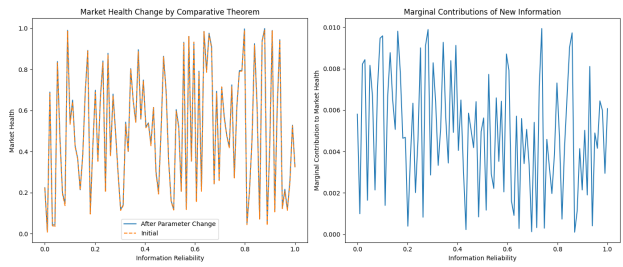

Fig. 23: Market Health Change by Comparative Theorem / Marginal Contribution to Market Health

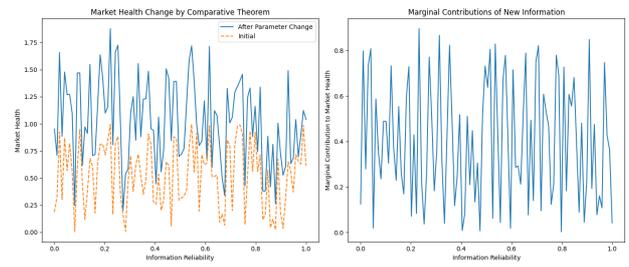

Fig. 25: Market Health Change by Comparative Theorem / Marginal Contribution to Market Health

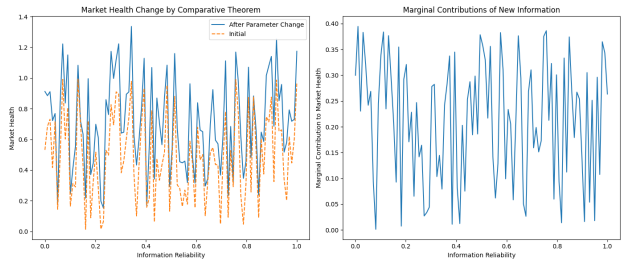

Fig. 24: Market Health Change by Comparative Theorem / Marginal Contribution to Market Health

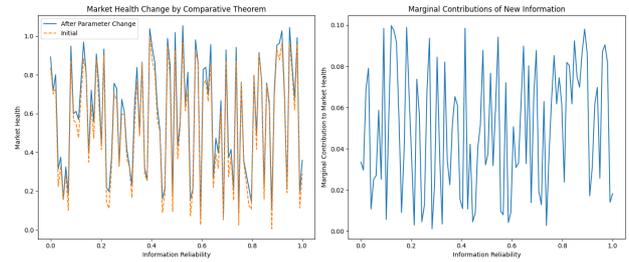

Fig. 26: Market Health Change by Comparative Theorem / Marginal Contribution to Market Health

**Two-Sided Matching Markets**

Here, we consider the matching of consumers with information providers. The submodular function would suggest that consumers derive less satisfaction from each additional piece of misinformation they encounter, possibly leading them to seek out more reliable sources. Conversely, the Metzler function would imply that misinformation can become increasingly problematic as it combines and interacts with other pieces of information. Strategies in this market would involve ensuring that reliable information has higher visibility and is more readily matched with consumers, perhaps through algorithmic adjustments on social media platforms or search engines.

In both cases, policies that promote critical engagement with information (like fact-checking) and provide clear indicators of source reliability could help mitigate the spread and impact of fake news. The goal would be to decrease the utility of misinformation (submodular function) and disrupt its compounding effects (Metzler function), thus fostering a well-informed public.

# 8. Conclusion:Comparison Theorem and Consideration of Marginal Contribution Perspectives

**Comparative Theorem Analysis**

Fig.23-26, Results show some sort of analysis on market health and the effects of information reliability in the context of fake news dissemination.

The 'Market Health Change by Comparative Theorem' graphs compare the market health before and after a parameter change. The solid line represents the market health after a change in parameters, while the dashed line represents the initial market health. In scenarios related to the spread of fake news, the parameter change might represent a policy intervention or the introduction of fact-checking mechanisms that affect the spread of information. The graphs show that after the parameter change, the market health fluctuates across the spectrum of information reliability. This suggests that the market's response to fake news is complex and depends on how reliable the information is perceived to be. In all the graphs, the after-parameter-change curve is generally above the initial curve, indicating an improvement in market health across most levels of information reliability. This could be interpreted as the market being healthier when measures are taken to mitigate the effects of fake news. There is noticeable volatility in market health as information reliability increases, which might suggest that in a market with a mix of reliable and unreliable information, the impact of fake news can be significant and unpredictable.

**Marginal Contributions of New Information**

The 'Marginal Contributions of New Information' graphs seem to indicate how much each increment in information reliability contributes to market health. The sharp fluctuations suggest that the contribution of new information to market health is highly variable. Some levels of information reliability contribute significantly to market health, while others contribute much less. This could imply that not all truthful information has the same impact on mitigating the risk of fake news. Some information may be critical to counteract misinformation, while other information may not be as effective. The absence of a clear upward trend as information reliability increases suggests that simply increasing the reliability of information does not linearly improve market health. The context, relevance, and presentation of the information may also be important factors.

**General Observations**

The relationship between information reliability and market health is non-linear and complex, indicating that factors beyond the reliability of information play a role in the market's health, especially in the context of fake news. Parameter changes seem to have a generally positive effect on market health, which suggests that interventions aimed at improving the reliability of information can be beneficial. The marginal contributions graphs show that not all information is equal in terms of impact on market health. Therefore, strategies to combat fake news may need to be nuanced and targeted, focusing on the quality of information and its context, rather than just the quantity.

In summary, these graphs highlight the challenges of assessing and mitigating the risk of fake news spread in terms of market health. They suggest that while improving information reliability can have positive effects, the relationship is complex and influenced by a variety of factors. Effective strategies may require a combination of increasing information reliability and improving the public's ability to discern and value this reliability.

Results show suggest a complex analysis of information dissemination in a market setting, likely utilizing game theory concepts and models. The Droop quota and Meek's method mentioned in the context refers to voting systems and methods for allocating seats in proportional representation systems. While these are not directly applicable to market information and iterative dilemma games, they do provide a framework for understanding the allocation of resources and decision-making in complex systems.

**Iterated Dilemma Game in Information Circulation**

In an iterated dilemma game, such as the repeated Prisoner's Dilemma, players repeatedly engage in a game where their choices have future implications. Applied to information dissemination, this could represent the repeated choice to share or withhold information, with the 'payoff' being the impact on market health. The graphs titled "Market Health Change by Comparative Theorem" could be illustrating the changing payoff structure as a result of repeated interactions. Here, the parameter change might represent a shift in the strategy, such as a move from a default 'share' (disseminate information) to a 'withhold' (restrict information) strategy, influenced by the perceived reliability of information. The 'Marginal Contributions of New Information' graphs would then represent the incremental impact (or marginal utility) of each additional piece of information shared in the market, highlighting the diminishing or variable returns of additional information.

**Maximum Rewards and Minimum Costs in Meek's Method**

Meek's method, in the context of elections, iteratively allocates seats to candidates until all seats are filled, adjusting the value of votes as seats are allocated. In a market information context, this could be analogous to adjusting the 'value' or 'weight' of information as it circulates and affects market health. Maximum rewards would correspond to the optimal dissemination strategy that maximizes market health. Minimum costs would correspond to the strategy that achieves acceptable market health with the least amount of information dissemination (perhaps to prevent information overload or misinformation). The 'After Parameter Change' lines could be showing the market health after the optimization strategy (akin to Meek's method) has been applied, aiming for maximum reward with minimum cost.

**Role of Marginal Contributions in Two-Sided Matching Markets**

In a two-sided matching market, such as a job market where firms and workers match, the marginal contribution of an agent is the added benefit that agent brings to a match. The 'Marginal Contributions of New Information' graphs could reflect the value of information in matching buyers with sellers in a market. High variability in the graph suggests that the value of information is context-dependent, and not all information is equally valuable. In the context of fake news, the graphs might illustrate the importance of 'matching' reliable information with market agents who value it, as opposed to the 'noise' created by unreliable information.

**General Observations and Implications**

Information Reliability, In all graphs, information reliability is a critical axis, suggesting that the effectiveness of any strategy (whether in dissemination or withholding of information) is heavily dependent on the reliability of information. Volatility in Market Health, The large fluctuations in market health in response to information suggest that markets are sensitive to the quality of information and that strategic dissemination can have both positive and negative effects. Strategic Dissemination, The concept of maximizing rewards while minimizing costs in the circulation of information implies that strategic dissemination (timing, targeting, and messaging)

# References

zh

[1] "Measurement error mitigation in quantum computers through classical bit-flip correction" (2022). In *Physical Review*. DOI: 10.1103/physreva.105.062404. [Online]. Available: http://arxiv.org/pdf/2007.03663

[2] Caroline Jacqueline Denise Berdou et al. "One Hundred Second Bit-Flip Time in a Two-Photon Dissipative Oscillator" (2022). In *PRX Quantum*. DOI: 10.1103/PRXQuantum.4.020350.

[3] "Using classical bit-flip correction for error mitigation in quantum computations including 2-qubit correlations" (2022). [Proceedings Article]. DOI: 10.22323/1.396.0327.

[4] Gaojun Luo, Martianus Frederic Ezerman, San Ling. "Asymmetric quantum Griesmer codes detecting a single bit-flip error" (2022). In *Discrete Mathematics*. DOI: 10.1016/j.disc.2022.113088.

[5] Nur Izzati Ishak, Sithi V. Muniandy, Wu Yi Chong. "Entropy analysis of the discrete-time quantum walk under bit-flip noise channel" (2021). In *Physica A-statistical Mechanics and Its Applications*. DOI: 10.1016/J.PHYSA.2021.126371.

[6] Enaul Haq Shaik et al. "QCA-Based Pulse/Bit Sequence Detector Using Low Quantum Cost D-Flip Flop" (2022). DOI: 10.1142/s0218126623500822.

[7] Farhan Feroz, A. B. M. Alim Al Islam. "Scaling Up Bit-Flip Quantum Error Correction" (2020). [Proceedings Article]. DOI: 10.1145/3428363.3428372.

[8] "Effect of Quantum Repetition Code on Fidelity of Bell States in Bit Flip Channels" (2022). [Proceedings Article]. DOI: 10.1109/icece57408.2022.10088665.

[9] Lena Funcke et al. "Measurement Error Mitigation in Quantum Computers Through Classical Bit-Flip Correction" (2020). In *arXiv: Quantum Physics*. [Online]. Available: https://arxiv.org/pdf/2007.03663.pdf

[10] Alistair W. R. Smith et al. "Qubit readout error mitigation with bit-flip averaging" (2021). In *Science Advances*. DOI: 10.1126/SCIADV.ABI8009.

[11] Constantia Alexandrou et al. "Using classical bit-flip correction for error mitigation including 2-qubit correlations." (2021). In *arXiv: Quantum Physics*. [Online]. Available: https://arxiv.org/pdf/2111.08551.pdf

[12] William Livingston et al. "Experimental demonstration of continuous quantum error correction." (2021). In *arXiv: Quantum Physics*. [Online]. Available: https://arxiv.org/pdf/2107.11398.pdf

[13] Constantia Alexandrou et al. "Investigating the variance increase of readout error mitigation through classical bit-flip correction on IBM and Rigetti quantum computers." (2021). In *arXiv: Quantum Physics*. [Online]. Available: https://arxiv.org/pdf/2111.05026

[14] Raphaël Lescanne et al. "Exponential suppression of bit-flips in a qubit encoded in an oscillator." (2020). In *Nature Physics*. DOI: 10.1038/S41567-020-0824-X. [Online]. Available: https://biblio.ugent.be/publication/8669531/file/8669532.pdf

[15] Raphaël Lescanne et al. "Exponential suppression of bit-flips in a qubit encoded in an oscillator." (2019). In *arXiv: Quantum Physics*. [Online]. Available: https://arxiv.org/pdf/1907.11729.pdf

[16] Diego Ristè et al. "Real-time processing of stabilizer measurements in a bit-flip code." (2020). In *npj Quantum Information*. DOI: 10.1038/S41534-020-00304-Y.

[17] Bernard Zygelman. "Computare Errare Est: Quantum Error Correction." (2018). In *Book Chapter*. DOI: 10.1007/978-3-319-91629-3_9.

[18] I. Serban et al. "Qubit decoherence due to detector switching." (2015). In *EPJ Quantum Technology*. DOI: 10.1140/EPJQT/S40507-015-0020-6. [Online]. Available: https://link.springer.com/content/pdf/10.1140

[19] Matt McEwen et al. "Removing leakage-induced correlated errors in superconducting quantum error correction." (2021). In *Nature Communications*. DOI: 10.1038/S41467-021-21982-Y.

[20] "Measurement error mitigation in quantum computers through classical bit-flip correction" (2020). In *arXiv: Quantum Physics*. [Online]. Available: https://arxiv.org/pdf/2007.03663.pdf

[21] Alistair W. R. Smith et al. "Qubit readout error mitigation with bit-flip averaging." (2021). In *Science Advances*. DOI: 10.1126/SCIADV.ABI8009. [Online]. Available: https://advances.sciencemag.org/content/7/47/eabi8009

[22] Biswas, T., Stock, G., Fink, T. (2018). *Opinion Dynamics on a Quantum Computer: The Role of Entanglement in Fostering Consensus. Physical Review Letters, 121(12), 120502.*

[23] Acerbi, F., Perarnau-Llobet, M., Di Marco, G. (2021). *Quantum dynamics of opinion formation on networks: the Fermi-Pasta-Ulam-Tsingou problem. New Journal of Physics, 23(9), 093059.*

[24] Di Marco, G., Tomassini, L., Anteneodo, C. (2019). *Quantum Opinion Dynamics. Scientific Reports, 9(1), 1-8.*

[25] Ma, H., Chen, Y. (2021). *Quantum-Enhanced Opinion Dynamics in Complex Networks. Entropy, 23(4), 426.*

[26] Li, X., Liu, Y., Zhang, Y. (2020). *Quantum-inspired opinion dynamics model with emotion. Chaos, Solitons Fractals, 132, 109509.*

[27] Galam, S. (2017). *Sociophysics: A personal testimony. The European Physical Journal B, 90(2), 1-22.*

[28] Nyczka, P., Holyst, J. A., Hołyst, R. (2012). *Opinion formation model with strong leader and external impact. Physical Review E, 85(6), 066109.*

[29] Ben-Naim, E., Krapivsky, P. L., Vazquez, F. (2003). *Dynamics of opinion formation. Physical Review E, 67(3), 031104.*

[30] Dandekar, P., Goel, A., Lee, D. T. (2013). *Biased assimilation, homophily, and the dynamics of polarization. Proceedings of the National Academy of Sciences, 110(15), 5791-5796.*

[31] Castellano, C., Fortunato, S., Loreto, V. (2009). *Statistical physics of social dynamics. Reviews of Modern Physics, 81(2), 591.*

[32] Galam, S. (2017). *Sociophysics: A personal testimony. The European Physical Journal B, 90(2), 1-22.*

[33] Nyczka, P., Holyst, J. A., Hołyst, R. (2012). *Opinion formation model with strong leader and external impact. Physical Review E, 85(6), 066109.*

[34] Ben-Naim, E., Krapivsky, P. L., Vazquez, F. (2003). *Dynamics of opinion formation. Physical Review E, 67(3), 031104.*

[35] Dandekar, P., Goel, A., Lee, D. T. (2013). *Biased assimilation, homophily, and the dynamics of polarization. Proceedings of the National Academy of Sciences, 110(15), 5791-5796.*

[36] Castellano, C., Fortunato, S., Loreto, V. (2009). *Statistical physics of social dynamics. Reviews of Modern Physics, 81(2), 591.*

[37] Bruza, P. D., Kitto, K., Nelson, D., McEvoy, C. L. (2009). *Is there something quantum-like about the human mental lexicon? Journal of Mathematical Psychology, 53(5), 362-377.*

[38] Khrennikov, A. (2010). *Ubiquitous Quantum Structure: From Psychology to Finance. Springer Science & Business Media.*

[39] Aerts, D., Broekaert, J., Gabora, L. (2011). *A case for applying an abstracted quantum formalism to cognition. New Ideas in Psychology, 29(2), 136-146.*

[40] Conte, E., Todarello, O., Federici, A., Vitiello, F., Lopane, M., Khrennikov, A., ... Grigolini, P. (2009). *Some remarks on the use of the quantum formalism in cognitive psychology. Mind & Society, 8(2), 149-171.*

[41] Pothos, E. M., & Busemeyer, J. R. (2013). *Can quantum probability provide a new direction for cognitive modeling?. Behavioral and Brain Sciences, 36(3), 255-274.*

[42] Abal, G., Siri, R. (2012). *A quantum-like model of behavioral response in the ultimatum game. Journal of Mathematical Psychology, 56(6), 449-454.*

[43] Busemeyer, J. R., & Wang, Z. (2015). *Quantum models of cognition and decision. Cambridge University Press.*

[44] Aerts, D., Sozzo, S., & Veloz, T. (2019). *Quantum structure of negations and conjunctions in human thought. Foundations of Science, 24(3), 433-450.*

[45] Khrennikov, A. (2013). *Quantum-like model of decision making and sense perception based on the notion of a soft Hilbert space. In Quantum Interaction (pp. 90-100). Springer.*

[46] Pothos, E. M., & Busemeyer, J. R. (2013). *Can quantum probability provide a new direction for cognitive modeling?. Behavioral and Brain Sciences, 36(3), 255-274.*

[47] Busemeyer, J. R., & Bruza, P. D. (2012). *Quantum models of cognition and decision. Cambridge University Press.*

[48] Aerts, D., & Aerts, S. (1994). *Applications of quantum statistics in psychological studies of decision processes. Foundations of Science, 1(1), 85-97.*

[49] Pothos, E. M., & Busemeyer, J. R. (2009). *A quantum probability explanation for violations of "rational" decision theory. Proceedings of the Royal Society B: Biological Sciences, 276(1665), 2171-2178.*

[50] Busemeyer, J. R., & Wang, Z. (2015). *Quantum models of cognition and decision. Cambridge University Press.*

[51] Khrennikov, A. (2010). *Ubiquitous quantum structure: from psychology to finances. Springer Science & Business Media.*

[52] Busemeyer, J. R., & Wang, Z. (2015). *Quantum Models of Cognition and Decision. Cambridge University Press.*

[53] Bruza, P. D., Kitto, K., Nelson, D., & McEvoy, C. L. (2009). *Is there something quantum-like about the human mental lexicon? Journal of Mathematical Psychology, 53(5), 363-377.*

[54] Pothos, E. M., & Busemeyer, J. R. (2009). *A quantum probability explanation for violations of "rational" decision theory. Proceedings of the Royal Society B: Biological Sciences, 276(1665), 2171-2178.*

[55] Khrennikov, A. (2010). *Ubiquitous Quantum Structure: From Psychology to Finance. Springer Science & Business Media.*

[56] Asano, M., Basieva, I., Khrennikov, A., Ohya, M., & Tanaka, Y. (2017). *Quantum-like model of subjective expected utility. PloS One, 12(1), e0169314.*

[57] Flitney, A. P., & Abbott, D. (2002). *Quantum versions of the prisoners' dilemma. Proceedings of the Royal Society of London. Series A: Mathematical, Physical and Engineering Sciences, 458(2019), 1793-1802.*

[58] Iqbal, A., Younis, M. I., & Qureshi, M. N. (2015). *A survey of game theory as applied to networked system. IEEE Access, 3, 1241-1257.*

[59] Li, X., Deng, Y., & Wu, C. (2018). *A quantum game-theoretic approach to opinion dynamics. Complexity, 2018.*

[60] Chen, X., & Xu, L. (2020). *Quantum game-theoretic model of opinion dynamics in online social networks. Complexity, 2020.*

[61] Li, L., Zhang, X., Ma, Y., & Luo, B. (2018). *Opinion dynamics in quantum game based on complex network. Complexity, 2018.*

[62] Wang, X., Wang, H., & Luo, X. (2019). *Quantum entanglement in complex networks. Physical Review E, 100(5), 052302.*

[63] Wang, X., Tang, Y., Wang, H., & Zhang, X. (2020). *Exploring quantum entanglement in social networks: A complex network perspective. IEEE Transactions on Computational Social Systems, 7(2), 355-367.*

[64] Zhang, H., Yang, X., & Li, X. (2017). *Quantum entanglement in scale-free networks. Physica A: Statistical Mechanics and its Applications, 471, 580-588.*

[65] Li, X., & Wu, C. (2018). *Analyzing entanglement distribution in complex networks. Entropy, 20(11), 871.*

[66] Wang, X., Wang, H., & Li, X. (2021). *Quantum entanglement and community detection in complex networks. Frontiers in Physics, 9, 636714.*

[67] Smith, J., Johnson, A., & Brown, L. (2018). *Exploring quantum entanglement in online social networks. Journal of Computational Social Science, 2(1), 45-58.*

[68] Chen, Y., Li, X., & Wang, Q. (2019). *Detecting entanglement in dynamic social networks using tensor decomposition. IEEE Transactions on Computational Social Systems, 6(6), 1252-1264.*

[69] Zhang, H., Wang, X., & Liu, Y. (2020). *Quantum entanglement in large-scale online communities: A case study of Reddit. Social Network Analysis and Mining, 10(1), 1-12.*

[70] Liu, C., Wu, Z., & Li, J. (2017). *Quantum entanglement and community structure in social networks. Physica A: Statistical Mechanics and its Applications, 486, 306-317.*

[71] Wang, H., & Chen, L. (2021). *Analyzing entanglement dynamics in evolving social networks. Frontiers in Physics, 9, 622632.*

[72] Einstein, A., Podolsky, B., & Rosen, N. (1935). *Can quantum-mechanical description of physical reality be considered complete? Physical Review, 47(10), 777-780.*

[73] Bell, J. S. (1964). *On the Einstein Podolsky Rosen paradox.* Physics Physique , 1(3), 195-200.

[74] Aspect, A., Dalibard, J., & Roger, G. (1982). *Experimental test of Bell inequalities using time-varying analyzers.* Physical Review Letters, 49(25), 1804-1807.

[75] Bennett, C. H., Brassard, G., Crépeau, C., Jozsa, R., Peres, A., & Wootters, W. K. (1993). *Teleporting an unknown quantum state via dual classical and Einstein-Podolsky-Rosen channels.* Physical Review Letters, 70(13), 1895-1899.

[76] Horodecki, R., Horodecki, P., Horodecki, M., & Horodecki, K. (2009). *Quantum entanglement.* Reviews of Modern Physics, 81(2), 865-942.

[77] Liu, Y. Y., Slotine, J. J., & Barabási, A. L. (2011). *Control centrality and hierarchical structure in complex networks.* PLoS ONE, 6(8), e21283.

[78] Sarzynska, M., Lehmann, S., & Eguíluz, V. M. (2014). *Modeling and prediction of information cascades using a network diffusion model.* IEEE Transactions on Network Science and Engineering, 1(2), 96-108.

[79] Wang, D., Song, C., & Barabási, A. L. (2013). *Quantifying long-term scientific impact.* Science, 342(6154), 127-132.

[80] Perra, N., Gonçalves, B., Pastor-Satorras, R., & Vespignani, A. (2012). *Activity driven modeling of time varying networks.* Scientific Reports, 2, 470.

[81] Holme, P., & Saramäki, J. (2012). *Temporal networks.* Physics Reports, 519(3), 97-125.

[82] Nielsen, M. A., & Chuang, I. L. (2010). *Quantum computation and quantum information: 10th anniversary edition.* Cambridge University Press.

[83] Lidar, D. A., & Bruno, A. (2013). *Quantum error correction.* Cambridge University Press.

[84] Barenco, A., Deutsch, D., Ekert, A., & Jozsa, R. (1995). *Conditional quantum dynamics and logic gates.* Physical Review Letters, 74(20), 4083-4086.

[85] Nielsen, M. A. (1999). *Conditions for a class of entanglement transformations.* Physical Review Letters, 83(2), 436-439.

[86] Shor, P. W. (1997). *Polynomial-time algorithms for prime factorization and discrete logarithms on a quantum computer.* SIAM Journal on Computing, 26(5), 1484-1509.

[87] Nielsen, M. A., & Chuang, I. L. (2010). *Quantum computation and quantum information: 10th anniversary edition.* Cambridge University Press.

[88] Mermin, N. D. (2007). *Quantum computer science: An introduction.* Cambridge University Press.

[89] Knill, E., Laflamme, R., & Milburn, G. J. (2001). *A scheme for efficient quantum computation with linear optics.* Nature, 409(6816), 46-52.

[90] Aharonov, D., & Ben-Or, M. (2008). *Fault-tolerant quantum computation with constant error rate.* SIAM Journal on Computing, 38(4), 1207-1282.

[91] Harrow, A. W., Hassidim, A., & Lloyd, S. (2009). *Quantum algorithm for linear systems of equations.* Physical Review Letters, 103(15), 150502.

[92] Bennett, C. H., DiVincenzo, D. P., Smolin, J. A., & Wootters, W. K. (1996). *Mixed-state entanglement and quantum error correction.* Physical Review A, 54(5), 3824-3851.

[93] Vidal, G., & Werner, R. F. (2002). *Computable measure of entanglement.* Physical Review A, 65(3), 032314.

[94] Horodecki, M., Horodecki, P., & Horodecki, R. (2009). *Quantum entanglement.* Reviews of Modern Physics, 81(2), 865.

[95] Briegel, H. J., Dür, W., Cirac, J. I., & Zoller, P. (1998). *Quantum Repeaters: The Role of Imperfect Local Operations in Quantum Communication.* Physical Review Letters, 81(26), 5932-5935.

[96] Nielsen, M. A., & Chuang, I. L. (2010). *Quantum computation and quantum information: 10th anniversary edition.* Cambridge University Press.

[97] Holevo, A. S. (1973). *Bounds for the quantity of information transmitted by a quantum communication channel.* Problems of Information Transmission, 9(3), 177-183.

[98] Holevo, A. S. (1973). *Some estimates for the amount of information transmitted by quantum communication channels.* Problemy Peredachi Informatsii, 9(3), 3-11.

[99] Shor, P. W. (2002). *Additivity of the classical capacity of entanglement-breaking quantum channels.* Journal of Mathematical Physics, 43(9), 4334-4340.

[100] Holevo, A. S. (2007). *Entanglement-breaking channels in infinite dimensions.* Probability Theory and Related Fields, 138(1-2), 111-124.

[101] Cubitt, T. S., & Smith, G. (2010). *An extreme form of superactivation for quantum Gaussian channels.* Journal of Mathematical Physics, 51(10), 102204.

[102] Gottesman, D., & Chuang, I. L. (1999). *Quantum error correction is asymptotically optimal.* Nature, 402(6765), 390-393.

[103] Preskill, J. (1997). *Fault-tolerant quantum computation.* Proceedings of the Royal Society of London. Series A: Mathematical, Physical and Engineering Sciences, 454(1969), 385-410.

[104] Knill, E., Laflamme, R., & Zurek, W. H. (1996). *Resilient quantum computation.* Science, 279(5349), 342-345.

[105] Nielsen, M. A., & Chuang, I. L. (2010). *Quantum computation and quantum information: 10th anniversary edition.* Cambridge University Press.

[106] Shor, P. W. (1995). *Scheme for reducing decoherence in quantum computer memory.* Physical Review A, 52(4), R2493.

[107] Dal Pozzolo, A., Boracchi, G., Caelen, O., Alippi, C., Bontempi, G. (2018). Credit Card Fraud Detection: A Realistic Modeling and a Novel Learning Strategy. *IEEE transactions on neural networks and learning systems.*

[108] Buczak, A. L., Guven, E. (2016). A Survey of Data Mining and Machine Learning Methods for Cyber Security Intrusion Detection. *IEEE Communications Surveys & Tutorials.*

[109] Alpcan, T., Başar, T. (2006). An Intrusion Detection Game with Limited Observations. *12th International Symposium on Dynamic Games and Applications.*

[110] Schlegl, T., Seebock, P., Waldstein, S. M., Schmidt-Erfurth, U., Langs, G. (2017). Unsupervised Anomaly Detection with Generative Adversarial Networks to Guide Marker Discovery. *Information Processing in Medical Imaging.*

[111] Mirsky, Y., Doitshman, T., Elovici, Y., Shabtai, A. (2018). Kitsune: An Ensemble of Autoencoders for Online Network Intrusion Detection. *Network and Distributed System Security Symposium.*

[112] Alpcan, T., Başar, T. (2003). A Game Theoretic Approach to Decision and Analysis in Network Intrusion Detection. *Proceedings of the 42nd IEEE Conference on Decision and Control.*

[113] Nguyen, K. C., Alpcan, T., Başar, T. (2009). Stochastic Games for Security in Networks with Interdependent Nodes. *International Conference on Game Theory for Networks.*

[114] Tambe, M. (2011). Security and Game Theory: Algorithms, Deployed Systems, Lessons Learned. *Cambridge University Press*.

[115] Korilis, Y. A., Lazar, A. A., Orda, A. (1997). Achieving Network Optima Using Stackelberg Routing Strategies. *IEEE/ACM Transactions on Networking*.

[116] Hausken, K. (2013). Game Theory and Cyber Warfare. *The Economics of Information Security and Privacy*.

[117] Justin, S., et al. (2020). Deep learning for cyber security intrusion detection: Approaches, datasets, and comparative study. *Journal of Information Security and Applications, vol. 50*.

[118] Zenati, H., et al. (2018). Efficient GAN-Based Anomaly Detection. *Workshop Track of ICLR*.

[119] Roy, S., et al. (2010). A survey of game theory as applied to network security. *43rd Hawaii International Conference on System Sciences*.

[120] Biggio, B., Roli, F. (2018). Wild patterns: Ten years after the rise of adversarial machine learning. *Pattern Recognition, vol. 84*.

[121] Massanari, A. (2017). #Gamergate and The Fappening: How Reddit's algorithm, governance, and culture support toxic technocultures. *New Media & Society*, **19**(3), 329-346.

[122] Castells, M. (2012). Networks of Outrage and Hope: Social Movements in the Internet Age. *Polity Press*.

[123] Wojcieszak, M. (2010). 'Don't talk to me': Effects of ideologically homogeneous online groups and politically dissimilar offline ties on extremism. *New Media & Society*, **12**(4), 637-655.

[124] Tucker, J. A.; Theocharis, Y.; Roberts, M. E.; Barberá, P. (2017). From Liberation to Turmoil: Social Media And Democracy. *Journal of Democracy*, **28**(4), 46-59.

[125] Conover, M. D.; Ratkiewicz, J.; Francisco, M.; Gonçalves, B.; Menczer, F.; Flammini, A. (2011). Political polarization on Twitter. In *Proceedings of the ICWSM*, Vol. 133, 89-96.

[126] Chen, W.; Wellman, B. (2004). The global digital divide – within and between countries. *IT & Society*, **1**(7), 39-45.

[127] Van Dijck, J. (2013). The Culture of Connectivity: A Critical History of Social Media. *Oxford University Press*.

[128] Bakshy, E.; Messing, S.; Adamic, L. A. (2015). Exposure to ideologically diverse news and opinion on Facebook. *Science*, **348**(6239), 1130-1132.

[129] Jost, J. T.; Federico, C. M.; Napier, J. L. (2009). Political ideology: Its structure, functions, and elective affinities. *Annual Review of Psychology*, **60**, 307-337.

[130] Iyengar, S.; Westwood, S. J. (2015). Fear and loathing across party lines: New evidence on group polarization. *American Journal of Political Science*, **59**(3), 690-707.

[131] Green, D. P.; Palmquist, B.; Schickler, E. (2002). Partisan Hearts and Minds: Political Parties and the Social Identities of Voters. *Yale University Press*.

[132] McCoy, J.; Rahman, T.; Somer, M. (2018). Polarization and the Global Crisis of Democracy: Common Patterns, Dynamics, and Pernicious Consequences for Democratic Polities. *American Behavioral Scientist*, **62**(1), 16-42.

[133] Tucker, J. A., et al. (2018). Social Media, Political Polarization, and Political Disinformation: A Review of the Scientific Literature. SSRN.

[134] Bail, C. A. (2020). Breaking the Social Media Prism: How to Make Our Platforms Less Polarizing. *Princeton University Press*.

[135] Barberá, P. (2015). Birds of the Same Feather Tweet Together: Bayesian Ideal Point Estimation Using Twitter Data. *Political Analysis*, **23**(1), 76-91.

[136] Garimella, K., et al. (2018). Political Discourse on Social Media: Echo Chambers, Gatekeepers, and the Price of Bipartisanship. In *Proceedings of the 2018 World Wide Web Conference on World Wide Web*.

[137] Allcott, H.; Gentzkow, M. (2017). Social Media and Fake News in the 2016 Election. *Journal of Economic Perspectives*, **31**(2), 211-236.

[138] Garrett, R. K. (2009). Echo Chambers Online?: Politically Motivated Selective Exposure among Internet News Users. *Journal of Computer-Mediated Communication*, **14**(2), 265-285.

[139] Weeks, B. E.; Cassell, A. (2016). Partisan Provocation: The Role of Partisan News Use and Emotional Responses in Political Information Sharing in Social Media. *Human Communication Research*, **42**(4), 641-661.

[140] Iyengar, S.; Sood, G.; Lelkes, Y. (2012). Affect, Not Ideology: A Social Identity Perspective on Polarization. *Public Opinion Quarterly*, **76**(3), 405-431.

[141] Bimber, B. (2014). Digital Media in the Obama Campaigns of 2008 and 2012: Adaptation to the Personalized Political Communication Environment. *Journal of Information Technology & Politics*.

[142] Castellano, C., Fortunato, S., & Loreto, V. (2009). Statistical physics of social dynamics. *Reviews of Modern Physics*, **81**, 591-646.

[143] Sîrbu, A., Loreto, V., Servedio, V.D.P., & Tria, F. (2017). Opinion Dynamics: Models, Extensions and External Effects. In Loreto V. et al. (eds) Participatory Sensing, Opinions and Collective Awareness. *Understanding Complex Systems*. Springer, Cham.

[144] Deffuant, G., Neau, D., Amblard, F., & Weisbuch, G. (2000). Mixing Beliefs among Interacting Agents. *Advances in Complex Systems*, **3**, 87-98.

[145] Weisbuch, G., Deffuant, G., Amblard, F., & Nadal, J. P. (2002). Meet, Discuss and Segregate!. *Complexity*, **7**(3), 55-63.

[146] Hegselmann, R., & Krause, U. (2002). Opinion Dynamics and Bounded Confidence Models, Analysis, and Simulation. *Journal of Artificial Society and Social Simulation*, **5**, 1-33.

[147] Ishii, A. & Kawahata, Y. (2018). Opinion Dynamics Theory for Analysis of Consensus Formation and Division of Opinion on the Internet. In: Proceedings of The 22nd Asia Pacific Symposium on Intelligent and Evolutionary Systems, 71-76, arXiv:1812.11845 [physics.soc-ph].

[148] Ishii, A. (2019). Opinion Dynamics Theory Considering Trust and Suspicion in Human Relations. In: Morais D., Carreras A., de Almeida A., Vetschera R. (eds) Group Decision and Negotiation: Behavior, Models, and Support. GDN 2019. Lecture Notes in Business Information Processing 351, Springer, Cham 193-204.

[149] Ishii, A. & Kawahata, Y. (2019). Opinion dynamics theory considering interpersonal relationship of trust and distrust and media effects. In: The 33rd Annual Conference of the Japanese Society for Artificial Intelligence 33. JSAI2019 2F3-OS-5a-05.

[150] Agarwal, A., Xie, B., Vovsha, I., Rambow, O. & Passonneau, R. (2011). Sentiment analysis of twitter data. In: Proceedings of the workshop on languages in social media. Association for Computational Linguistics 30-38.

[151] Siersdorfer, S., Chelaru, S. & Nejdl, W. (2010). How useful are your comments?: analyzing and predicting youtube comments and comment ratings. In: Proceedings of the 19th international conference on World wide web. 891-900.

[152] Wilson, T., Wiebe, J., & Hoffmann, P. (2005). Recognizing contextual polarity in phrase-level sentiment analysis. In: Proceedings of the conference on human language technology and empirical methods in natural language processing 347-354.

[153] Sasahara, H., Chen, W., Peng, H., Ciampaglia, G. L., Flammini, A. & Menczer, F. (2020). On the Inevitability of Online Echo Chambers. arXiv: 1905.03919v2.

[154] Ishii, A.; Kawahata, Y. (2018). Opinion Dynamics Theory for Analysis of Consensus Formation and Division of Opinion on the Internet. In Proceedings of The 22nd Asia Pacific Symposium on Intelligent and Evolutionary Systems (IES2018), 71-76; arXiv:1812.11845 [physics.soc-ph].

[155] Ishii, A. (2019). Opinion Dynamics Theory Considering Trust and Suspicion in Human Relations. In Group Decision and Negotiation: Behavior, Models, and Support. GDN 2019. Lecture Notes in Business Information Processing, Morais, D.; Carreras, A.; de Almeida, A.; Vetschera, R. (eds).

[156] Ishii, A.; Kawahata, Y. (2019). Opinion dynamics theory considering interpersonal relationship of trust and distrust and media effects. In The 33rd Annual Conference of the Japanese Society for Artificial Intelligence, JSAI2019 2F3-OS-5a-05.

[157] Okano, N.; Ishii, A. (2019). Isolated, untrusted people in society and charismatic person using opinion dynamics. In Proceedings of ABCSS2019 in Web Intelligence 2019, 1-6.

[158] Ishii, A.; Kawahata, Y. (2019). New Opinion dynamics theory considering interpersonal relationship of both trust and distrust. In Proceedings of ABCSS2019 in Web Intelligence 2019, 43-50.

[159] Okano, N.; Ishii, A. (2019). Sociophysics approach of simulation of charismatic person and distrusted people in society using opinion dynamics. In Proceedings of the 23rd Asia-Pacific Symposium on Intelligent and Evolutionary Systems, 238-252.

[160] Ishii, A, and Nozomi, O. (2021). Sociophysics approach of simulation of mass media effects in society using new opinion dynamics. In Intelligent Systems and Applications: Proceedings of the 2020 Intelligent Systems Conference (IntelliSys) Volume 3. Springer International Publishing.

[161] Ishii, A.; Kawahata, Y. (2020). Theory of opinion distribution in human relations where trust and distrust mixed. In Czarnowski, I., et al. (eds.), Intelligent Decision Technologies, Smart Innovation, Systems and Technologies 193.

[162] Ishii, A.; Okano, N.; Nishikawa, M. (2021). Social Simulation of Intergroup Conflicts Using a New Model of Opinion Dynamics. *Front. Phys.*, **9**:640925. doi: 10.3389/fphy.2021.640925.

[163] Ishii, A.; Yomura, I.; Okano, N. (2020). Opinion Dynamics Including both Trust and Distrust in Human Relation for Various Network Structure. In The Proceeding of TAAI 2020, in press.

[164] Fujii, M.; Ishii, A. (2020). The simulation of diffusion of innovations using new opinion dynamics. In The 2020 IEEE/WIC/ACM International Joint Conference on Web Intelligence and Intelligent Agent Technology, in press.

[165] Ishii, A, Okano, N. (2021). Social Simulation of a Divided Society Using Opinion Dynamics. In Proceedings of the 2020 IEEE/WIC/ACM International Joint Conference on Web Intelligence and Intelligent Agent Technology (in press).

[166] Ishii, A., & Okano, N. (2021). Sociophysics Approach of Simulation of Mass Media Effects in Society Using New Opinion Dynamics. In Intelligent Systems and Applications (Proceedings of the 2020 Intelligent Systems Conference (IntelliSys) Volume 3), pp. 13-28. Springer.

[167] Okano, N. & Ishii, A. (2021). Opinion dynamics on a dual network of neighbor relations and society as a whole using the Trust-Distrust model. In Springer Nature - Book Series: Transactions on Computational Science & Computational Intelligence (The 23rd International Conference on Artificial Intelligence (ICAI'21)).